# Vortex-dynamical Interpretation of Anti-phase and In-phase Flickering of Dual Buoyant Diffusion Flames

Tao Yang[1], Xi Xia[1,2], and Peng Zhang[1,*]

[1]*Department of Mechanical Engineering,*

*The Hong Kong Polytechnic University, Hung Hom, Kowloon, Hong Kong*

[2]*School of Mechanical Engineering,*

*Shanghai Jiao Tong University, Minhang, Shanghai, P. R. China*

Anti-phase and in-phase flickering modes of dual buoyant diffusion flames were numerically investigated and theoretically analyzed in this study. Inspired by the flickering mechanism of a single buoyant diffusion flame, for which the deformation, stretching, or even pinch-off of the flame surface result from the formation and evolution of the toroidal vortices, we attempted to understand the anti-phase and in-phase flickering of dual buoyant diffusion flames from the perspective of vortex dynamics. The interaction between the inner-side shear layers of the two flames was identified to be responsible for the different flickering modes. Specifically, the transition between anti-phase and in-phase flickering modes can be predicted by a unified regime nomogram of the normalized flickering frequency versus a characteristic Reynolds number, which accounts for the viscous effect on vorticity diffusion between the two inner-side shear layers. Physically, the transition of the vortical structures from symmetric (in-phase) to staggered (anti-phase) in a dual-flame system can be interpreted as being similar to the mechanism causing flow transition in the wake of a bluff body and forming the Karman vortex street.

* Corresponding author
  E-mail pengzhang.zhang@polyu.edu.hk.
  Fax: (852)23654703   Tel: (852)27666664

# I. INTRODUCTION

The interaction of multiple flames is a crucial problem that concerns flame stability and fire safety in industrial and environmental applications [1-4]. Probably due to its intrinsic unsteadiness, this problem has not been sufficiently understood, even for the simplest scenario—the interaction of two flames [5-9]. In fact, the unsteadiness could originate from a single flame, for example the "flickering" or "puffing" of a buoyant diffusion flame [10-21]. Flickering is a periodic flame oscillation phenomenon. The flame is elongated vertically and contracted horizontally to form a "neck", the upper portion (above the neck) of the flame is pinched off, the lower portion (below the neck) of the flame retracts, and then the next cycle starts [14].

For single diffusion flames without or with small initial velocity, buoyancy is the main source that induces the flame to evolve into a self-sustained oscillation mode [12,13,22,23]. The occurrence of flame oscillation phenomena, such as flame pinch-off and flame flickering, can be determined by a Rayleigh number criterion and a Froude number criterion, respectively proposed by Carpio *et al.* [14] and Moreno-Boza *et al.* [15]. Recently, a scaling relation for the frequency of flickering diffusion flames has been analytically obtained by Xia and Zhang [24] based on vortex-dynamical analysis. Physically, flame flickering has been attributed to the dynamics of toroidal vortices [13], as illustrated in Fig. 1(a). The lower-density hot gas around the non-premixed flame sheet is accelerated upwardly by buoyancy, leading to the formation of a thin shear layer (or equivalently a vortex sheet) immediately outside the flame sheet at Stage 1. Starting from the flame anchoring base, the shear layer rolls up into a toroidal vortex at Stage 2, which grows and convects upwardly, gradually deforms the flame to form a thin neck at Stage 3, and eventually breaks the neck and causes the flame pinch-off at Stage 4. The detached flame pocket is then lifted in a flow driven by the detached toroidal vortex [13,14], while new vortex sheet starts to grow outside the attached flame sheet to form the toroidal vortex of the next cycle.

Flickering also exists in multi-flame systems. Recent experimental studies [25-29] have reported that for candle or jet flames in pairs or arrays the flickering of each individual flame spontaneously synchronizes with each other, rendering special synchronized flickering modes. According to Kitahata *et al.* [25], the interaction between two side-by-side oscillating flames exhibits two distinct modes, in-phase and anti-phase, depending on the distance between the flames. In the in-phase flickering, the two flames stretch and pinch off in a simultaneous manner as if there were only one flickering flame. In the anti-phase flickering, each flame experiences pinch-off alternatively in a "seesaw" manner. Forrester [27] experimentally identified an initial-arch-bow-initial "worship" oscillation mode for four candles arranged within a closed loop. Okamoto *et al.* [28] reported four distinct synchronized flickering modes for three coupled candle flames in a triangular array. Despite the complexity of these multi-flame systems, the crucial component of them is the interaction of a dual-flame system, which is largely affected by the gap distance. Researchers proposed hypotheses to explain the coupling interaction by using the Hopf bifurcation theory based on heat transfer [25] and the conceptualized vortex-based conjecture [28]. However, no evidence of the flow field was provided to substantiate these hypotheses.

As discussed above, a single buoyant diffusion flame flickers under the periodic flow induced by the formation and detachment of the toroidal vortices. Two identical flames with sufficiently large separation distance are expected to have independent flickering processes with the same frequency. As the flames are closer to one another, each of them undergoes a disturbance from the other, which breaks the symmetry of the established flow pattern around each flame. Naturally, the toroidal vortex together with the flow would become asymmetric, which could in turn cause the anti-phase flame flickering. As the flames are so close to nearly merge with each other, it is also intuitive to perceive the in-phase flickering of the flames, which could have a different flickering frequency as the result of changed flame shape and flow field. Apparently, the above conjectures need to be verified and the underlying physics must be elucidated.

The present study aims to theoretically analyze the in-phase and anti-phase flickering modes of a dual-flame system from the vortex dynamics perspective. The existing literature does not provide enough details about the flow field information required by the analysis, which will be obtained in the study by means of numerical simulation. By analyzing the interaction between the shear layers of the two flames, we revealed the fundamental mechanism that governs the flickering modes.

**II. NUMERICAL METHODS**

In the present study, we were mainly concerned with the flickering mechanism of the pool flame duet as the result of their interaction, so two interacting laminar flames from square-shaped pools were considered without loss of generality, as illustrated in Fig. 1(b). The two pool bases are of the same dimensions with a fixed edge length, $d$, while the gap distance, $l$, in-between was adjusted to obtain different interactions between the flames. It is recognized that the flickering frequency is subject to more factors such as the geometry of the pools, the boundary conditions, the fuel thermochemistry, the transport of gases, and probably the turbulent flow. The present study does not however aim to precisely predict the frequency by considering the secondary effects caused by the many factors, but it focuses on using the simplified flame system to unravel the controlling vortex-dynamical mechanism underneath the different flickering modes.

The present work employed the computational fluid dynamics method implemented in the widely-used open-source code, Fire Dynamics Simulator (FDS) [30], to solve the unsteady, three-dimensional, incompressible (variable-density) flow with chemical heat release, the governing equations for which are expressed as

$$\frac{\partial}{\partial t}(\rho) + \nabla \cdot (\rho \boldsymbol{u}) = \dot{m}_F''', \tag{1}$$

$$\frac{\partial}{\partial t}(\rho Y_\alpha) + \nabla \cdot (\rho Y_\alpha \boldsymbol{u}) = \nabla \cdot (\rho D_\alpha \nabla Y_\alpha) + \dot{m}_\alpha''' + \dot{m}_F''', \tag{2}$$

$$\frac{\partial}{\partial t}(\rho \boldsymbol{u}) + \nabla \cdot (\rho \boldsymbol{u}\boldsymbol{u}) = -\nabla \tilde{p} - \nabla \cdot \tau + (\rho - \rho_0)\boldsymbol{g}, \tag{3}$$

$$\frac{\partial}{\partial t}(\rho h_s) + \nabla \cdot (\rho h_s \boldsymbol{u}) = \frac{D\bar{p}}{Dt} + \dot{q}''' - \nabla \cdot \dot{\boldsymbol{q}}'', \tag{4}$$

$$\rho = \frac{\bar{p}\overline{W}}{RT}, \tag{5}$$

where $\rho$ is the density, $\boldsymbol{u}$ the velocity vector, and $\dot{m}_F'''$ the mass production rate per unit volume of fuel by evaporation; $Y_\alpha$ and $D_\alpha$ are the mass fraction and diffusion coefficient of species $\alpha$, respectively; $\dot{m}_\alpha'''$ is the mass production rate per unit volume of species $\alpha$ by chemical reactions; $\tilde{p}$ is the pressure perturbation, $\tau$ the viscous stress, $\rho_0$ the background density, $\boldsymbol{g}$ the gravity vector, $h_s$ the sensible enthalpy under low Mach number approximation, $\bar{p}$ the back pressure, $\dot{q}'''$ the heat release per unit volume, $\dot{\boldsymbol{q}}''$ the heat flux vector, $\overline{W}$ the molecular weight of gas mixture, $R$ the universal gas constant, and $T$ the temperature. In the past decade, FDS has been proven to be suitable for capturing unsteady and dynamic processes in fire-driven flows [31-38]. Xin *et al.* [33] quantitatively reproduced the velocity field of a 1m-diameter methane pool fire using FDS. Hietaniemi *et al.* [34] verified that for heptane pool fires of various diameters the simulated burning rates agree well with experiments. The radiation solver of FDS was testified in modeling small pool fires by Hostikka *et al.* [35].

For simulation set up, the gray exterior surfaces in Fig. 1(b) are applied the open boundary condition, which allows gases to flow into or out of the computational domain depending on the local pressure gradient. Specifically, the Poisson solver requires a Dirichlet condition for the quantity, $\Omega = \tilde{p}/\rho + \boldsymbol{u} \cdot \boldsymbol{u}/2$, which means that $\Omega$ should remain constant along a streamline. Thus, for incoming flow $\Omega_\text{ext}$ at the boundary equals that at infinity of the ambient flow, whereas for outgoing flow $\Omega_\text{ext}$ is set to the value of its adjacent interior grid cell. It is noted that similar Dirichlet condition setup is also adopted at an open boundary for other fluid properties, such as temperature and species mass fractions. The blue surface denotes the liquid fuel of heptane, whose vaporization rate is a function of the liquid temperature and the vapor concentration according to the Clausius-Clapeyron relation [39]. The base of each pool is enforced with the solid-wall boundary condition, which is impermeable, non-slip, and adiabatic. The discretization is performed using a kinetic-energy-conserving central difference schemes while the time is advanced by an explicit second-order predictor/corrector scheme [30]. The combustion is considered as the mixing-limited, infinitely fast reaction of lumped species [30], which is a good approximation for the present non-premixed flames far from extinction. Although radiation is unlikely to be significant in the present small-scale flames, the radiation transport equation, which is routinely included in the applications of FDS, is solved using finite volume method without modeling the soot formation. A typical computational domain has a dimension of 16 cm × 16 cm × 24 cm with a grid resolution of 160 × 160 × 240, which was determined based on the domain and grid independent studies in Section III. Both the computational domain and mesh are adjusted accordingly as the gap distance enlarges.

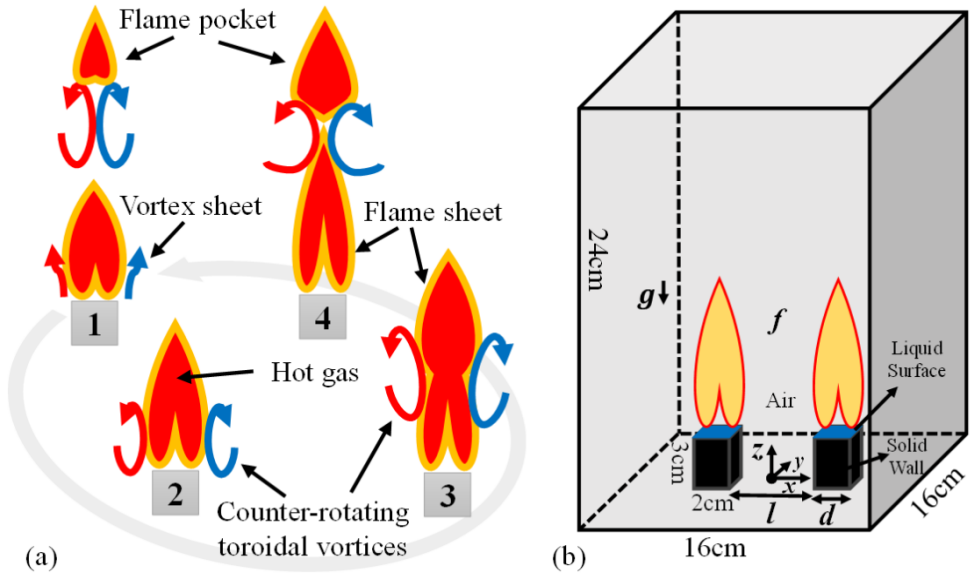

FIG. 1. (a) Illustration of the periodic flickering process of a single pool flame. (b) Schematic of the computational domain and definition of main parameters in the dual flame system.

## III. FLCKERING OF A SINGLE POOL FLAME

The flickering of a representative single heptane flame was simulated to validate our numerical method. The Reynold number, $Re = Ud/\nu_A$, is about 550, where the characteristic velocity $U$ is estimated as the convective velocity of the vortex scaled on $\sqrt{gd}$ with $g = 9.8 \text{ m/s}^2$; $\nu_A = 1.6 \times 10^{-5} \text{ m}^2/\text{s}$ is the kinematic viscosity of air at room temperature. The evolution of the simulated temperature and vorticity contours during a time period (denoted by $\Delta t = 1/f$, where $f$ is the flickering frequency) are presented in Fig. 2(a), which qualitatively reproduces the flickering phenomena of Fig. 1(a). It is clearly seen that the flame pinches off between $0.6\Delta t$ and $0.8\Delta t$ as a result of vorticity accumulation inside the toroidal vortex illustrated by the spiral streamlines, which induces high velocity near the center and causes the deformation and then rapture of the flame neck. After the flame pinch-off, the upper portion of the separated flame is convected downstream with the toroidal vortex. The right column of Fig. 2(a) presents the temperature distributions in three transverse planes at $z/d$ = 2.5, 3.5, and 5.5, justifying that the rectangular pool shape only affects the flame up to about $1.5d$ downstream of the fuel inlet.

Table 1. The grid- and domain- independence studies

| Cases | Domain (mm × mm × mm) | Mesh | $f_0$ (Hz) | $f_1$ (Hz) |
|---|---|---|---|---|
| Case 1 | 80 × 80 × 240 | 160 × 160 × 480 | 10.0 | 20.0 |
| Case 2 | 80 × 80 × 240 | 80 × 80 × 240 | 9.8 | 19.5 |
| Case 3 | 160 × 160 × 240 | 160 × 160 × 240 | 9.9 | 19.8 |
| Case 4 | 240 × 240 × 240 | 240 × 240 × 240 | 9.9 | 19.9 |

To facilitate the following parametric study, a balance between computational accuracy and efficiency is desirable. The grid- and domain-independence studies were conducted by comparing the characteristic frequencies of four cases of different domains and mesh sizes listed in Table 1. These frequencies were obtained by applying the Fast Fourier Transform (FFT) to analyze the mass burning rate of fuel for 10 s at a sampling rate of 1000 Hz so that the maximum and minimum of burning rates were converged, an example for which is shown in Fig. 2(b). The results indicate that the burning rate varies periodically with time, corresponding to a primary frequency of $f_0$ which is also the flickering frequency. The flickering flame has a sub-harmonic frequency $f_1$, which is twice of the primary frequency. Physically, the harmonics is closely related to the breakup of the main toroidal vortices in the downstream, which was found to be sensitive to environmental disturbances in previous experiments of laminar diffusion flames [15,17,19]. Based on the result of Table 1, the setting of Case 3, which features a domain of 16 cm × 16 cm × 24 cm and a structured, uniform, and staggered grid of $160 \times 160 \times 240$, was adopted as the standard domain and mesh for all single flame simulations.

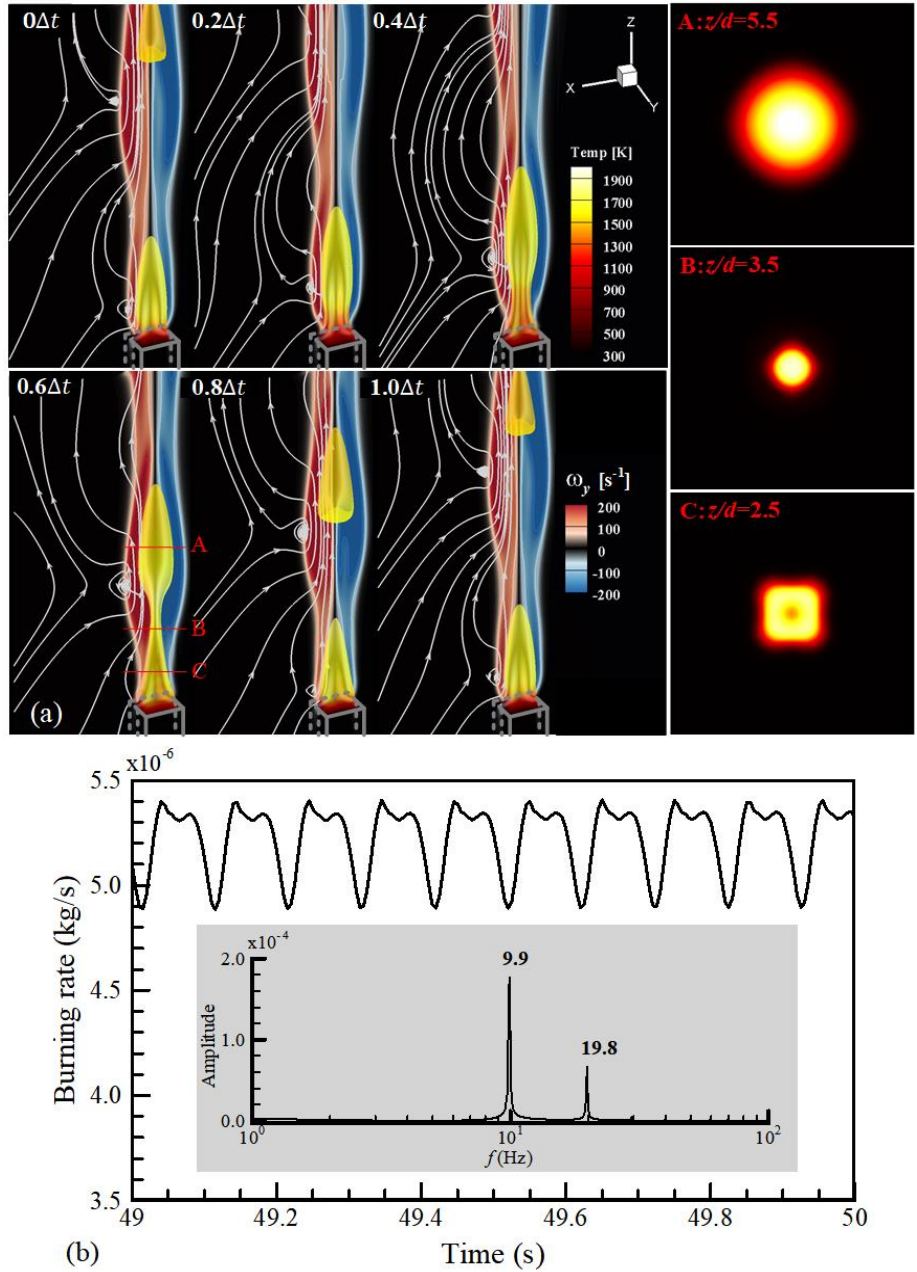

FIG. 2. (a) Visualization of time-varying single pool flame with $d = 20$ mm based on the flame sheet colored by temperature contour and vorticity on $X - Z$ plane, $Y = 0$. (b) The burning rate in time domain and frequency domain (Case 3).

Recently, we adopted vortex-dynamical theory to analytically derive the flickering frequency of a buoyancy-dominated diffusion flame as $f = 0.48\sqrt{g/d}$ [24], which recovers the prominent scaling relation obtained from previous experimental studies [40,41]. This scaling indicates that the flickering frequency depends largely on gravity and burner size, but less on the nozzle geometry, the fuel type or the initial fuel velocity. This allows us to conduct numerical experiment of flickering flames and adjust the flame parameters to check the performance of the current numerical approach. Here, different single pool flames with varying $g$ (4.9, 9.8 and 19.6 m/s$^2$) and $d$ (10, 15, 20, 30 mm) were simulated, and the simulation results in Fig. 3 agree well with previous experiments [42-45] and the scaling relation [24,40], verifying that the present numerical methods are suitable for studying the flickering of pool flames.

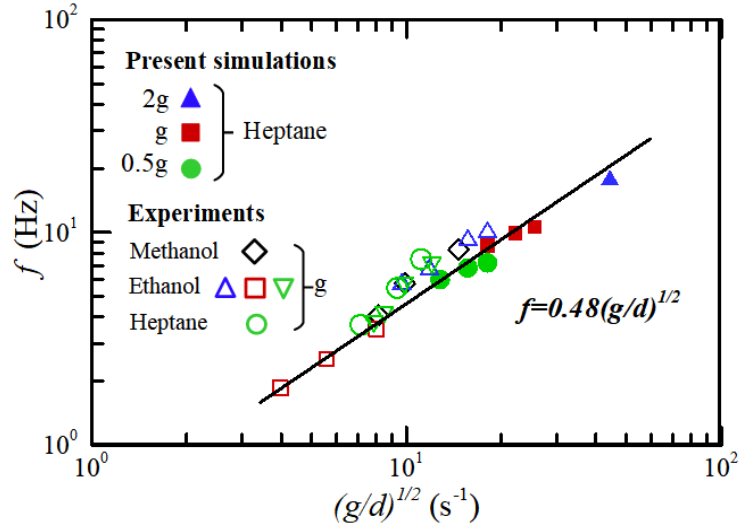

FIG. 3. Validation of single pool flames with the scaling law [40] and experiments obtained by Schönbucher *et al.* [42] (◊), Maynard [43] (△), Baum and McCaffrey [44] (□), and Fang *et al.* [45] (▽,○).

To study the essence of dual flickering flames, it is necessary to summarize the existing perspectives on interpreting the unsteady characteristics of a single flickering flame. In the present study, we have the perspective that the flickering flame is physically caused by the dynamics of vortices, as confirmed by numerous experimental studies, and modeled the flickering frequency by calculating the formation and detachment of the main toroidal vortex that is synchronized with the dynamics of the flame [24]. In another perspective, Moreno-Boza *et al.* [15] and others [11,46] interpreted the flickering as a hydrodynamic global instability. As such, the dynamics of the vortices are not modelled physically but treated as a perturbation, which could grow and propagate in the form of instability waves. It should be noted that both perspectives have been adopted extensively to study flow instability problems, among which a prominent one is the vortex shedding in the wake of a bluff body that renders the Karman vortex street. Admittedly, in terms of stability analysis, the introduction of the concepts of global/convective instabilities [47] enabled the successful application of the Landau equation [48] in modeling the onset of vortex shedding and led to the identification of the critical

Reynolds number [49], at which the stability mode transitions from a fixed point to a limit-cycle oscillation physically known as the vortex street. However, this more recent progress does not negate the earlier contributions by vortex dynamicists, including von Karman [50] and Saffman [51,52], who modeled the fully-developed vortex street with an infinite array of staggered counter-rotating vortices, resulting in the predictions of the characteristic Strouhal number based on linear stability theory. Therefore, the stability-analysis perspective based on Landau equation predicts the onset of the instability, whereas the vortex-dynamics perspective dictates the shedding frequency by assuming established instability of the vortex street. Apparently, these perspectives focus on the different stages of the vortex street development and are complementary to each other.

In the present study, there are two key advantages in applying the vortex-dynamics approach to study flame flickering. The first is the clear physical picture, which might be hidden in a pure mathematical approach simply treating a vortical structure as a wave disturbance. Second, it handles the unsteadiness by nature since vortices are the essence of flow motions. However, the stability-analysis approach generally requires starting from a stable base flow; thereby its applicability is limited to only a certain (often very short) period of time after the onset of instability.

## IV. PHENOMENA OF ANTI- AND IN- PHASE FLICKERING OF DUAL POOL FLAMES

### A. Anti-phase flickering mode

For two identical pool flames with a considerable separation distance in between, Kitahata *et al.* [25] reported the existence of an anti-phase synchronized flickering, and their experimental images during one period of flickering, are shown in Fig. 4(a). It is seen that each flame still stretches and shrinks in a periodic manner, while the flickering of the dual-flame system displays a "seesaw" pattern. It is evident that the flame pinch-off occurs at $0 \sim 0.2\Delta t$ for the left flame and at $0.4\Delta t \sim 0.6\Delta t$ for the right flame, and that the two flames are in opposite phases to each other. Thus, this oscillation mode is defined as "anti-phase" in this study.

The simulated heat release field is presented in Fig. 4(b) for comparison. It should be clarified that this simulation does not replicate the exact conditions of the original experiment [25] and, therefore, is not a numerical reproduction. This is because the experiment was conducted by using candle flames, with each "single" flame created by bundling three 6mm-diameter small candles together, so that the large "single" flame could pinch off [25]. Consequently, the differences in both the fuel inflow condition and the geometric shape of the fire base could cause the simulated pool flames to differ from the experimental candle flames. Nevertheless, the simulation still qualitatively captures the pinch-off of each individual flame as well as the "seesaw" oscillation pattern of the dual-flame system. Furthermore, it can be observed from the pinch-off behaviors of the simulation that the phase difference between the two flames is also around half a period, quantitatively verifying the establishment of the same anti-phase flickering.

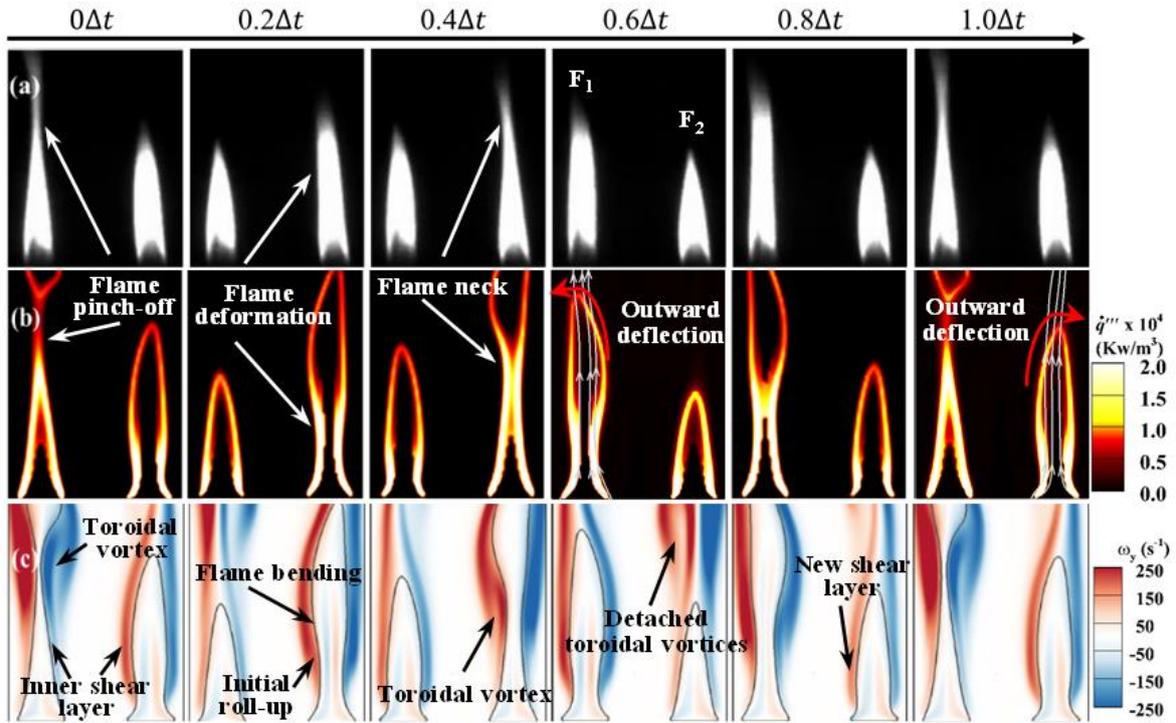

FIG. 4. Snapshots of flickering flames in the anti-phase mode. (a) The brightness from the experiment [25] based on two sets of three candles, (b) the heat release and (c) the vorticity from the present simulation based on two pool flames with $d = 20$ mm and $l = 30$ mm.

Fig. 4(c) presents the vorticity contours corresponding to the flame snapshots of Fig. 4(b). The black contour is obtained based on the stoichiometric mixture fraction to represent the flame sheet, which agrees well with the flame sheet visualized by heat release in Fig. 4(b). It can be observed that the flame sheet is accompanied by strong vortex flows, which is similar to the flame-vortex configuration of a single flame. Specifically, the flame deformation is related to the initial roll-up and growth of the shear layer, the flame neck appears with the growth of the toroidal vortex, and the flame pinch-off is synchronized with the detachment of the toroidal vortex. However, the difference is that the symmetry associated with each individual flame is disrupted as the flame and the shear layer seem to bend outwards, away from each other, as illustrated by the red arrows in Fig. 4(b). This can be understood by that each single flame (denoted $F_1$) is perturbed by the shear layers of the other flame (denoted $F_2$), whose net effect is to induce a velocity field indicated by the outwardly deflected streamline and to cause the outward deflection of $F_1$. It should be noted that, since the inner-side shear layer of $F_2$ is closer to $F_1$ than the outer-side shear layer, $F_1$ is dominated by the inner-side shear layer so that the induced velocity is deflected outwardly.

**B. In-phase flickering mode**

As the distance between the two flames becomes sufficiently small, the flickering mode changes dramatically that the two flames exhibit an in-phase synchronized oscillation. Fig. 5(a) shows the experimental snapshots of Forrester [27], who conducted a similar experiment to that of Kitahata *et al.* [25] using single candle instead of bundled candles. The simulation results of heat release based on pair pool flames are illustrated in Fig. 5(b). Both results show that the two flames flicker along with each other symmetrically as one flame, which is termed "in-phase" in the present study. The agreement in flame deformation, necking, and pinch-off between experiment and simulation is good in general, except the shape of the flame pocket during flame pinch-off (from $0.4\Delta t$ to $0.5\Delta t$). The discrepancy is likely to reflect the different setup between the experiment and the present simulation, for example, the different fuel and inflow conditions between candle flame and pool flame.

To investigate this in-phase mode of flame flickering, we again refer to the vorticity evolution during one flickering period, as shown in Fig. 5(c). It is interesting to note that the dynamics of the flames is still synchronized with the evolution of the vortical structures. Specifically, the flame deformation, necking, and pinch-off are respectively coupled with the initial roll-up, growth, and detachment of the toroidal vortex. However, the vortex sheet configuration for each individual flame displays a distinct asymmetric feature that the shear layer only rolls up into a vortex on the outer side with the inner side remains a rather linear shear layer, as observed at $0.4\Delta t$ and $0.5\Delta t$ in Fig. 5(c). Furthermore, the contour of each flame tends to deflect inward towards each other as indicated by the red arrows in Fig. 5(b), opposite to the outward flame deflection in the anti-phase mode. This again can be explained by the velocity-induction mechanism discussed in Section IV.A; however, the inward deflection of the current in-phase mode shares the same velocity direction as that induced by the outer-side shear layer of the other flame, as illustrated by the inwardly deflected streamline through the flame tip at $1\Delta t$ of Fig. 5(b). As the outer-side shear layer is farther away, it turns out that the outer-side shear layer together with the rolled-up vortex must have a significantly larger strength in order to dominate over the inner-side shear layer. These observations suggest that there is an imbalance of vortex strength between the outer and inner shear layers, the detailed mechanism for which will be investigated further in the next section.

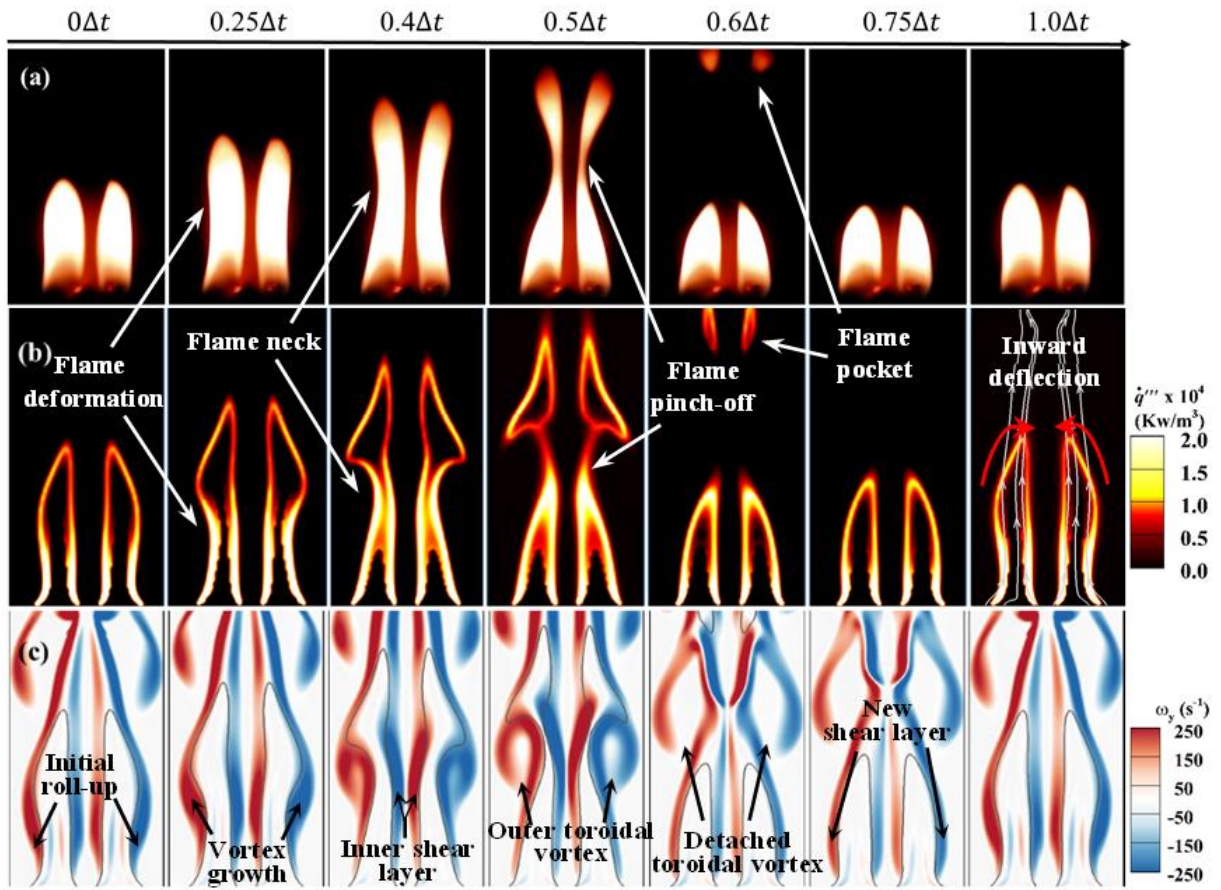

FIG. 5. Snapshots of flickering flames in the in-phase mode. (a) The brightness from the experiment [27] based on two candles, (b) the heat release and (c) the vorticity from present simulation based on two pool flames with $d = 20$ mm and $l = 10$ mm.

## V. VORTEX-DYNAMICAL INTERPRETATION OF DIFFERENT FLICKERING MODES

### A. Vorticity transport mechanism

To further understand the basic mechanisms for the anti- and in- phase flickering modes of the dual flame system, we recall that the flickering of a single flame is attributed to the periodicity of the toroidal vortices [13,14,40]. Our recent work [24] has demonstrated that the flickering frequency can be rigorously derived from the periodic generation and detachment of the toroidal vortices by applying vortex dynamical principles. A key mechanism here is the formation of the toroidal vortices, which is closely related to the generation and transportation of vorticity inside the flame-induced shear layer and is governed by the vorticity transport equation for incompressible flow,

$$\frac{D\boldsymbol{\omega}}{Dt} = \frac{\partial \boldsymbol{\omega}}{\partial t} + \underbrace{(\boldsymbol{u} \cdot \nabla)\boldsymbol{\omega}}_{(A)} = \underbrace{(\boldsymbol{\omega} \cdot \nabla)\boldsymbol{u}}_{(B)} + \underbrace{\frac{1}{\rho^2}(\nabla\rho \times \nabla\tilde{p})}_{(C)} + \underbrace{\frac{\rho_0}{\rho^2}(\nabla\rho \times \boldsymbol{g})}_{(D)} + \underbrace{\nu_A \nabla^2 \boldsymbol{\omega}}_{(E)}, \tag{6}$$

where the term (B) only accounts for the vorticity change associated with the stretching and tilting effects of the flow, the term (E) is a diffusion term causing the redistribution of vorticity, and both terms are not source of vorticity. Vorticity generation is attributed to the baroclinic term (C) and the gravitational term (D), both of which entail the density gradient caused by the heat release of combustion. For buoyancy-driven flames, $|\nabla\tilde{p}| \ll \rho_0 g$ so the dominant contributor to vorticity generation is term (D). From the observation of Section IV, it is evident that the roll-up of toroidal vortices also presents in both flickering modes of two flames, indicating that the mechanism of the growth and evolution of the shear layers of a single flame might be extended to understand the case of two flames.

The mechanism responsible for the anti-phase mode can be further interpreted based on the above discussion. For each individual flame in Fig. 4(c), the growth of the vortex sheet and the roll-up of the toroidal vortex resemble those in a single flame, so the main vorticity source is still term (D) in Eq. 6. The effect of flame $F_2$ on $F_1$ can be treated as a small perturbation, which from a vortex dynamics perspective results in an asymmetric vorticity convection term (A) in Eq. 6. This corresponds to an induced velocity field and slightly deforms flame $F_1$. One would expect this effect to be marginal;

however, the interactions between the shear layers as well as the toroidal vortices of the two flames clearly form a staggered distribution of the vortical structures, which is translated into the half-period phase difference between the two flames. As will be further discussed in Section VI, such staggered vortex configuration is typical in canonical fluid dynamics phenomena [53,54], for example the vortex street of a wake, and is usually the outcome of the development of flow instabilities.

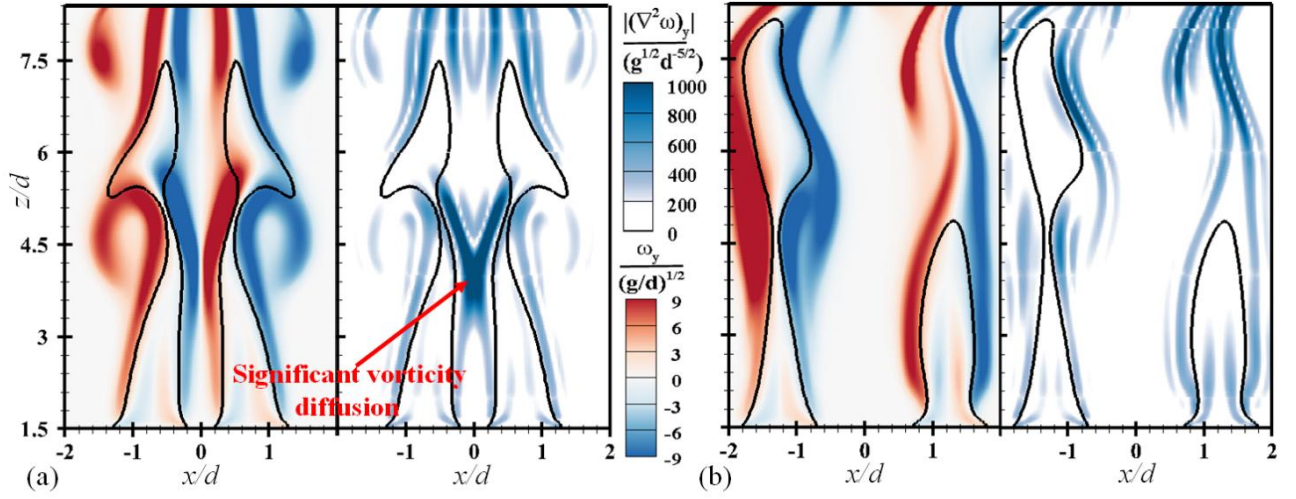

FIG. 6. Comparison between instantaneous dimensionless vorticity, $\omega_y/(g/d)^{1/2}$, and dimensionless Laplacian of vorticity, $|\nabla^2 \omega_y|/(g^{1/2} d^{-5/2})$, in (a) in-phase and (b) anti-phase flickering modes.

The analysis from Section IV.B points to the existence of a mechanism that causes the inner-side shear layers of the two flames to be weaker than the outer-side shear layers. From the vorticity contours of Fig. 4 and Fig. 5, it seems that the inner-side shear layers only become staggered without being diminished in the anti-phase mode, whereas the vortex strengths of the inner-side shear layers are notably weakened compared with the outer-side in the in-phase mode. This effect is quantitatively verified in Fig. 6, which shows that vorticity diffusion between the two inner-side shear layers causes a significant vorticity annihilation in the center plane of the two flames the in-phase mode. Mathematically, the vorticity diffusion corresponds to term (E) of Eq. 6, which intensifies with increasing vorticity gradient, and thus decreasing distance between the vortices. This explains the strong vorticity annihilation observed in the in-phase mode when the two inner-side shear layers approach each other. As will be further analyzed in the next section, this diffusion mechanism has a decisive effect on the transition between the two different flickering modes. It is noted that as the gap between the two flames decreases the density gradient across the inner-side shear layers could become smaller owing to thermal diffusion. As a result, it could also contribute to the reduction of vorticity of the inner shear layers. However, we believe that this heat diffusion effect should be secondary to the vorticity diffusion mechanism in determining the flickering modes, for which a detailed argument and analysis is provided in the Appendix.

**B. Helicity analysis**

Before moving on to the next section, we shall make a few notes to justify the approach of applying two-dimensional (2D) flow analysis to the current three-dimensional (3D) problem. First, the interaction between two flames mainly happens in the plane connecting them, away from which the interaction causes secondary effects. Furthermore, the instabilities and interactions of the vortices, which may enhance the 3D effects, mainly occur in the far field and thereby does not affect the laminar flow in the near field. This can be demonstrated by the helicity plots in Fig. 7, where the helicity density is given by $h = \boldsymbol{u} \cdot \boldsymbol{\omega}$ [55] with $\boldsymbol{\omega}$ being the vorticity vector. This definition means that helicity is generated by the "non-orthogonality" of the vorticity and velocity vectors. Physically, helicity in fluid dynamics is a measure of the knottedness and/or linkage of the vortex lines, and its volumetric integral inside an ideal Euler flow is an invariant given no vorticity crossing at the boundary. In the current context, the helicity may serve as a reference for the intensity of 3D interactions between the vortical structures. It is evident from Fig. 7(a) that in both in-phase and anti-phase flickering modes vortex lines located approximately below $z/d = 5$, where the streamlines become spiral and flame pinch-off occurs, are in regular 2D shape and black color (low $h$ value), indicating relative weak knottedness or linkage of the laminar vortical structures in the near field. However, the vortex lines in the downstream become twisted and irregular, and dyed in high $h$-value colors, which implies strong 3D interactions between the vortices. This can be understood by that the flow is mainly dominated by 2D vortex interactions in the near field where the shear layers remain laminar so that the out-plane velocity component is insignificant; whereas the vortex interactions become 3D in the far field when the shear layers evolve into more complex vortical structures and are entangled with each other. This understanding can be further demonstrated in Fig. 7(b), which shows the maximum and minimum of the areal integral of helicity density, $dH/dz = \iint h\,dxdy\ (x \geqq 0, y \geqq 0)$, within a flickering period at different $z$ locations. It is seen that in both flickering modes the helicity integral in the near field at $z < 5d$ is almost negligible while it increases notably in the downstream after the flame pinch-off location. Furthermore, the larger amplitude of helicity integral of the anti-phase mode compared with the in-phase mode clearly indicates a stronger 3D effect in the downstream of the anti-phase mode, reflecting a more unstable nature of flickering.

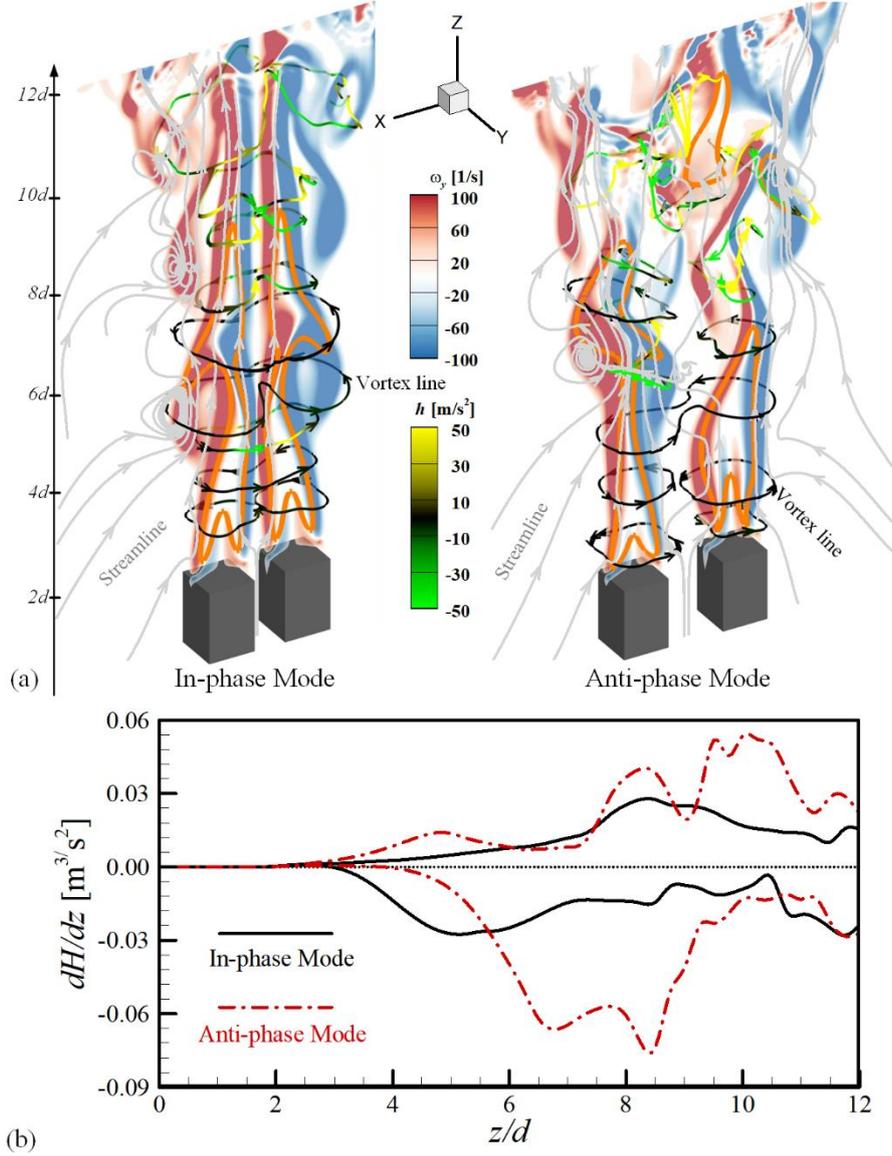

FIG. 7. (a) The distribution of vorticity in the $X - Z$ plane for in-phase (left) and anti-phase (right) flickering modes. The flame sheets are represented by the orange contours. The streamlines are denoted by the gray lines, and the vortex lines are colored by the helicity density, $h = \boldsymbol{u} \cdot \boldsymbol{\omega}$. (b) The area integral of helicity density, $dH/dz = \iint h\,dxdy\ (x \geq 0, y \geq 0)$, along the $z$ direction for in-phase (black solid lines) and anti-phase (red dash-dot lines) flickering modes. The two curves of each mode correspond to the upper and lower limits of $dH/dz$ during an entire flickering period.

**VI. CRITERION FOR FLICKERING MODE TRANSITION**

Hereto, we have attributed the anti-phase and in-phase flickering modes of a dual flame system to the interaction and evolution of vortical structures in the flow field. This contrasts the view proposed by Kitahata *et al.* [25] that the synchronization is owing to radiation, which was combined with the symmetric Hopf bifurcation theory [28] to predict the different flickering modes. The validity of Kitahata *et al.*'s model was later questioned in their other work [29] as it completely ignored the fluid dynamical effect. A more reasonable conjecture was given by Nakamura *et al.* [26] that the viscosity suppresses the hydrodynamic disturbance and leads to the mode transition of double flickering jet flames. It is worth mentioning that Dange *et al.* [56] recently provided a direct experimental support to our hypothesis [57] that the mode transition is determined by the interactions between the buoyancy-induced vortices. In this section, we shall focus on quantifying the interaction between the inner-side shear layers of the two different flames to reveal the nature of the transition between the two flickering modes.

**A. Dimensional analysis**

We first plot the non-dimensional frequency of double flickering flames, $f/f_s$, against $l/d$ in Fig. 8(a), where $f_s$ is the frequency of a single flickering flame, based on Kitahata *et al.*'s experiment of candle flames [25], Nakamura *et al.*'s experiment of jet flame [26], and the current simulations of pool flames. For Kitahata *et al.*'s experiment, the effective diameter of the bundled candles is estimated to be 13mm, which is the diameter of a virtual cylinder wrapping the candles. It is seen that in all cases the in-phase and anti-phase modes are distinguished by a frequency "jump", where $f/f_s$ increases from below 1 to above 1 within small ranges of $l/d$ as marked by the line segments in the figure. This frequency variation trend agrees well with existing literature [25,26,58]. However, the transition regions for different cases are scattered in Fig. 8(a), indicating that the synchronized flickering mode of two flames is not dictated by the non-dimensional gap distance $l/d$.

Next, we perform dimensional analysis to study the parameters governing the transition between the different flickering modes. Based on the phenomenal findings and physical understandings from Sections IV and V, the frequency $f$ of double flickering flames is a function of the pool size $d$, the gravity $g$, the gap distance $l$, and the ambient air viscosity $\nu_A$ as

$$f = F(d, g, l, \nu_A), \tag{7}$$

which yields

$$\frac{f}{f_s} \sim \frac{f}{\sqrt{g/d}} = G\left(\frac{l}{d}, \frac{\sqrt{gd^3}}{\nu_A}\right) = G(\alpha, Gr^{1/2}), \tag{8}$$

where $\alpha = l/d$ and $Gr = gd^3/\nu_A^2$ is the Grashof number [39] measuring the relative effects between buoyance-induced convective acceleration and viscous forces. A further observation of Fig. 8(a) suggests two limiting cases. As $\alpha \to 0$ by letting $l \to 0$, $f/f_s$ tends to approach a constant around 0.9, which agrees with 0.866 calculated according to the formula given by Tang $et\ al.$ [59]. As $\alpha \to \infty$ by letting $l \to \infty$, $f/f_s$ approaches 1 suggesting the desynchronization between the two flames [25,28]. It is evident that in both limiting cases $f/f_s$ depends more on $\alpha$ rather than $Gr^{1/2}$.

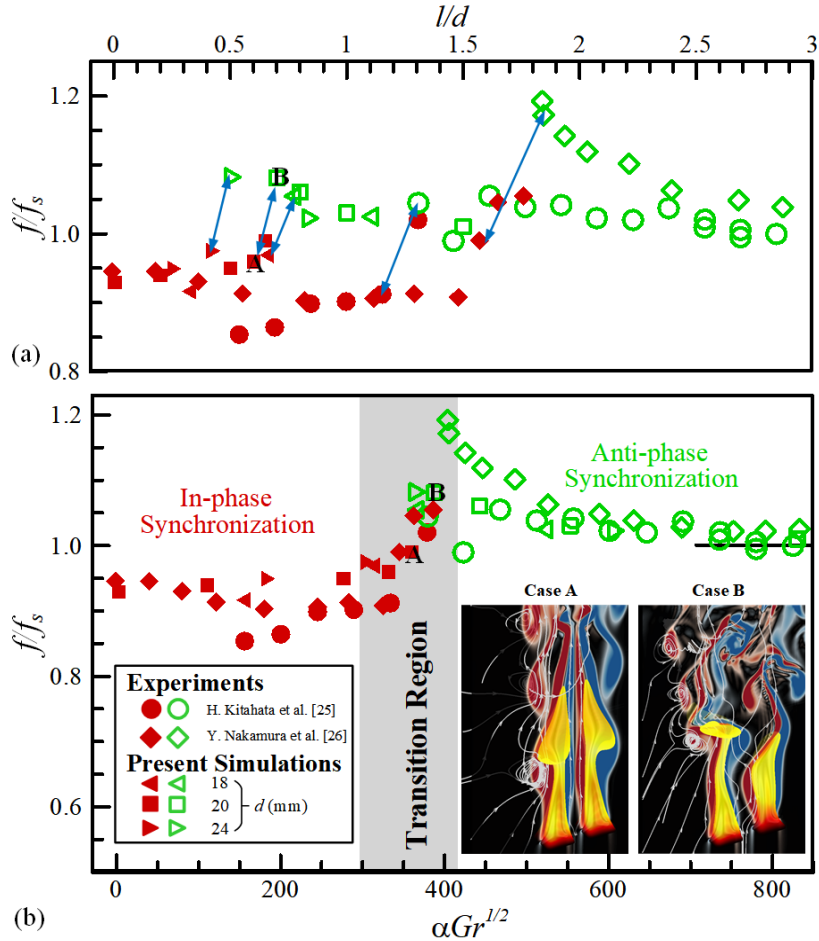

FIG. 8. Comparison of (a) normalized $f/f_s$ at different $l/d$ and (b) normalized $f/f_0$ at different $\alpha Gr^{1/2}$. The red solid and green empty symbols correspond to in-phase and anti-phase flickering modes, respectively. The line segments mark the transition between the two distinct modes, for instance, the flame and flow patterns of which correspond to Case A and Case B, respectively. The circle symbols are from Kitahata *et al.*'s experiment [25], the diamond ones from Nakamura *et al.*'s experiment [26], and the square and triangle symbols from present simulation.

The similarities in these asymptotic behaviors and the "jump" pertaining to the transition between the different flickering modes inspire us to find a sole parameter that dictates the variation of $f/f_s$ from unity to another constant. According to previous analysis, such a parameter, if existed, must be a combination of both $\alpha$ and $Gr^{1/2}$, and it should share the same asymptotic behaviors with $\alpha$ in the two limits, $\alpha \to 0$ and $\alpha \to \infty$. As a result, we found $\alpha Gr^{1/2}$ to be the desired parameter which can be used to collapse the $f/f_s$ data of different cases onto a single trend line as shown in Fig. 8(b), and the different transition regions in Fig. 8(a) are shrunk to a narrow band of the range $300 < \alpha Gr^{1/2} < 420$. This verifies the message from Eq. (8) that $\alpha$ and $Gr^{1/2}$ together govern the non-dimensional flickering frequency of two flames, including the transition between the two different flickering modes. Considering the vast differences between the candle, jet and pool flames, this result also suggests that the synchronized flickering is less sensitive to the nature or geometry of the flame, and $\alpha Gr^{1/2}$ could serve as a unified criterion to predict the flickering transition of a general dual-flame system.

## B. Physical interpretation

It is noted that the product of $\alpha$ and $Gr^{1/2}$ should not be seen as an arbitrary combination, but an integral parameter identified experimentally based on the hypothesis of a single criterion for the flickering mode transition. In the following we demonstrate that its underlying mechanism is consistent with physical interpretation. For this purpose, $\alpha Gr^{1/2}$ can be rewritten as

$$\alpha Gr^{1/2} = \frac{\sqrt{gd}\, l}{\nu_A} \sim \frac{Ul}{\nu_A} = Re_A, \tag{9}$$

where $U \sim \sqrt{gd}$ is the characteristic velocity. Eq. (9) implies that the transition between the anti- and in- phase flickering modes is dictated by the Reynolds number $Re_A$, which is defined based on the two inner-side shear layers of the gap flow. This Reynolds number mechanism can be understood by comparing the interaction between the inner-side shear layers of the two flames to that of the wake of a bluff body, e.g., the flow around a cylinder [53]. One similarity is the directions of the vorticities in the two counter-rotating shear layers. Furthermore, both flows involve a transition of the shear-layer configuration from symmetric to staggered, which is also known as the Karman vortex street [54]. From the analysis of Section V, we are more convinced to make this analogy because the transition is closely related to viscosity in both cases. For the canonical flow around a cylinder [53], this flow transition is predicted by a Reynolds number, $Re$, which quantifies the effect of inertia relative to viscosity. It was found that the transition from symmetric vortical structures to staggered vortex street usually happens around a critical Reynolds number from several decades to a hundred. In the current study of double flames, the transition from in-phase to anti-phase has been identified to occur at a few hundreds of $Re_A$. This confirms that the transition of the two different flickering modes of a dual-flame system is governed by the gap flow between the inner-side shear layers, which resembles the transition mechanism from stable to unstable in the development of a von Karman vortex street.

To further support the analogy between the gap flow of the two flames and the wake flow, we calculated the spacing ratio, $l/h^+$, for the staggered vortical structures in the simulated anti-phase flickering flames, where $h^+$ denotes the inter-core spacing of the main vortices in each inner-side shear layer. This definition is consistent with that for the classical Karman vortex street, and a sample calculation is shown in Fig. 9. Here, the vortex core, denoted by black cross, is identified by calculating the second invariant Q [60] and then finding the local maximum along each branch of the shear layers as demonstrated in Fig. 9(b). It is noted that this calculation should be performed in the near field where the staggered vortical structures are prominent and not much interfered by the 3D effects in the far field. In practice, $h^+$ is the averaged vertical height between two vortex cores and obtained by averaging the instantaneous values over nine phases of a period as shown in Fig. 10. As a result, the phase-averaged spacing ratio $l/h^+$ for different anti-phase flickering flames is estimated to vary between 0.20 and 0.33. This is in reasonable agreement with the theoretical value 0.28 suggested by von Karman [54] and the experimental values around 0.2 measured from different types of wakes [61], especially considering that the strength of the buoyancy-induced shear layers are growing gradually along the flame and the vortex pattern in the near field is not yet fully developed.

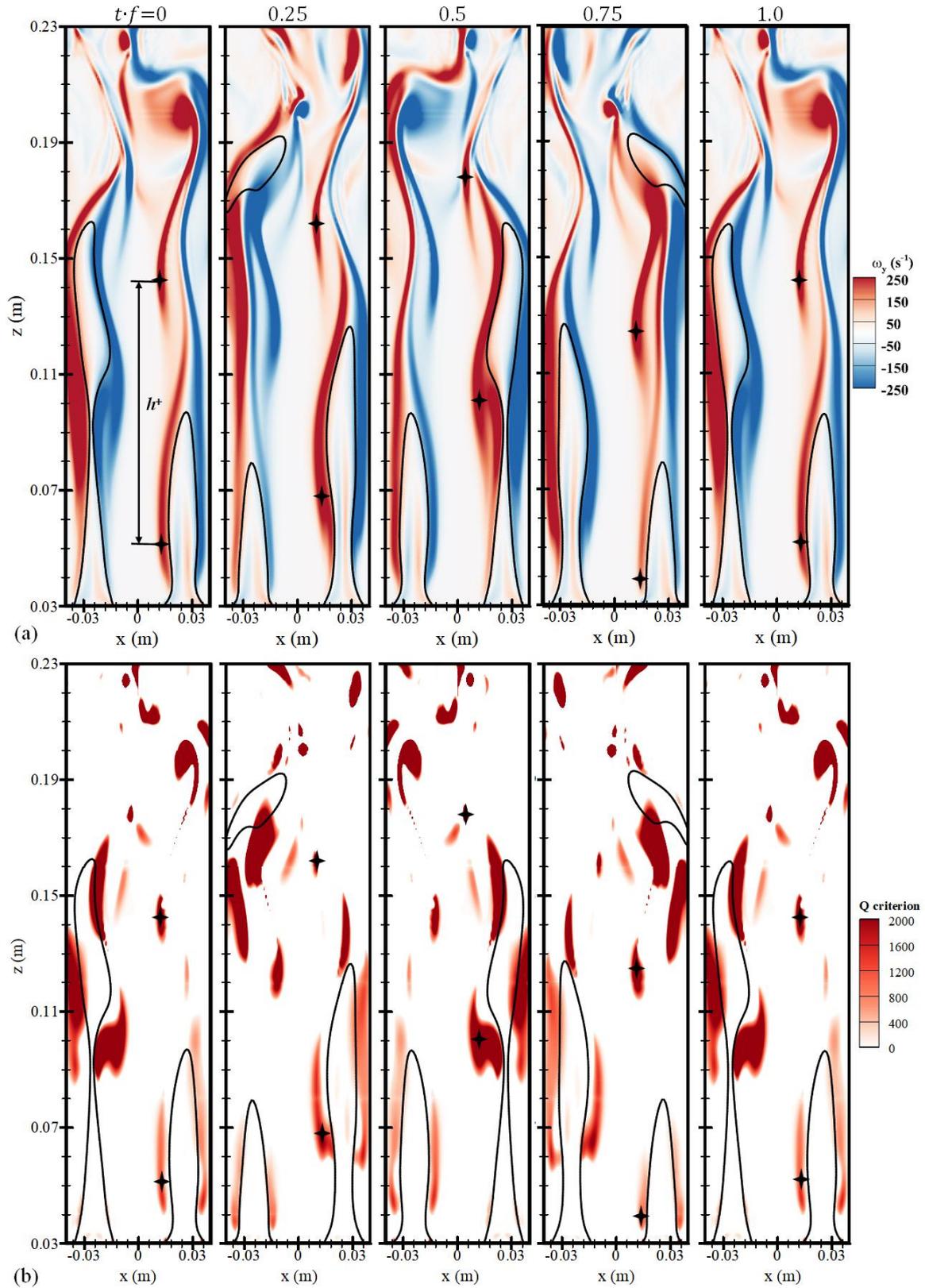

FIG. 9. Five instantaneous (a) vorticity and (b) Q-criterion contours of anti-phase flickering flames ($d = 2$ cm) with 30 mm gap. The black star denotes the position of a vortex core identified by the local maximum Q-criterion [60]. $h^+$ is the vertical height between two adjacent vortex cores.

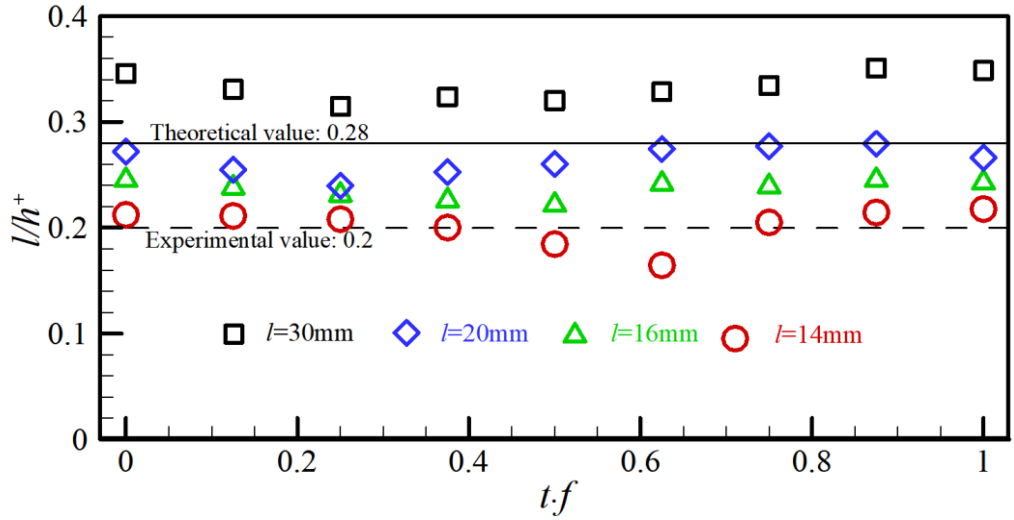

FIG.10. The spacing ratio $l/h^+$ during a period for four anti-phase flickering modes, which are marked by black □ ($l = 30$ mm), blue ◊ ($l = 20$ mm), green △ ($l = 16$ mm), and red ○ ($l = 14$ mm), respectively. The solid line is the theoretical value of 0.28 given by Von Karman [50] and the dashed line represents the experimental value around 0.2 [61].

**C. Further validation and discussion**

So far, this study has pointed to the conclusion that the mode transition of two flames is primarily dominated by fluid mechanics, through the viscous diffusion of vorticity. For further validation, Fig. 11 shows the mode transitions for the double pool flames of diameter $d = 20$ mm, at three different gap distances, 10 mm, 14 mm, and 20 mm, by numerically tuning the viscosity. We can observe that the mode transitions from in-phase to anti-phase as the viscosity coefficient decreases from $0.50\nu_A$, $1.10\nu_A$, and $1.35\nu_A$ to $0.45\nu_A$, $1.05\nu_A$, and $1.30\nu_A$, respectively. This directly verifies our proposed mechanism that the viscosity plays an essential role in the interaction between the inner shear layers of the flames. It should be noted that in previous studies all mode transitions of double flames were achieved through adjusting the gap distance, whereas our simulation shows the first evidence that the transition is possible by changing the viscosity alone. Another interesting observation is the flame interaction mode of Fig. 11(e), which is termed the "amplitude death" mode [56] meaning no flickering. While future investigation is necessary, we believe this mode happens when the viscous diffusion of each single flame is so strong that it prevents the shear layer from growing and rolling up into toroidal vortex, meanwhile the interaction between the inner shear layers is still in the small $Re_A$ regime so that the flame system maintains a symmetric configuration. Therefore, this is considered as a special in-phase mode in the present study.

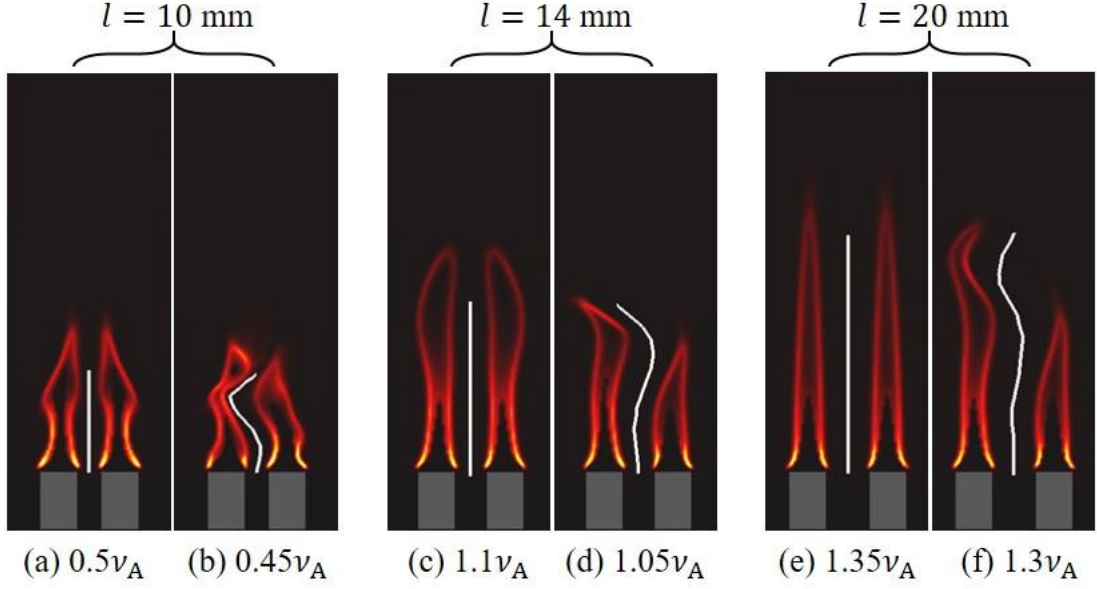

FIG.11. The contours of heat release for different simulation cases, showing the mode transition at different gap distances by changing the viscosity alone. The white lines are iso-contours of zero vorticity, along which the ambient flow temperature is estimated.

To further verify the mode transition criterion, we apply Eq. (9) to calculate the $Re_A$ transition ranges for the three gap distances of Fig. 11 to be 554~616, 353~370, and 410~426, respectively. The $Re_A$ ranges for the gap distances of 14 mm and 20 mm are in reasonable agreement with the predicted $Re_A$ range of 300~420 from Fig. 8; however, the $Re_A$ range for $l = 10$ mm is notably above the prediction. We believe that this mismatch should be attributed to the increased viscosity associated with the increased temperature of the gap flow as the gap shrinks. It can be corrected by modifying Eq. (9) as

$$Re_E = \frac{\sqrt{gd}\, l}{\nu_E}, \tag{10}$$

where $\nu_E$ is the effective kinematic viscosity estimated from the Sutherland's law as

$$\frac{\nu_E}{\nu_{ref}} = \left(\frac{T_E}{T_{ref}}\right)^{5/2} \frac{T_{ref} + S}{T_E + S}, \tag{11}$$

with $T_{ref} = 298$ K being the room temperature, $\nu_{ref}$ the reference viscosity at $T_{ref}$, $T_E$ the effective temperature, and $S = 110.4$ K for air. Note that the reduction of the density terms in Eq. (11) is based on the ideal gas law. Since the temperature of the gap flow is highly unsteady, $T_E$ is only evaluated approximately based on spatial and temporal averaging. For each case in Fig. 11, we first identify the white line corresponding to the iso-contour of zero vorticity within the gap. Thus, the time-averaged temperature along the white line, which is plotted in Fig. 12, is considered to be that of the ambient flow outside the inner shear layers. Furthermore, we believe that the flickering mode is decided by the vortex interaction upstream of the pinch-off location, $z_{po}$, which is marked by the dashed lines in Fig. 12. Consequently, $T_E$ for each case can be calculated by averaging the corresponding temperature curve of Fig. 12 for $z$ up to $z_{po}$. This allows the estimation of $Re_E$ based on Eqs. (10) and (11), and the results are presented in Table 2. We can see that the agreement between the $Re_E$ transition ranges and the range of 300~420 in Fig. 8 is significantly improved, especially for the $l = 10$ mm case.

Table 2. Reynolds number for mode transition by changing viscosity

| Gap distance | $l = 10$ mm | | $l = 14$ mm | | $l = 20$ mm | |
| --- | --- | --- | --- | --- | --- | --- |
| Mode | In-phase | Anti-phase | In-phase | Anti-phase | In-phase | Anti-phase |
| $\nu_{ref}$ | $0.50\nu_A$ | $0.45\nu_A$ | $1.10\nu_A$ | $1.05\nu_A$ | $1.35\nu_A$ | $1.30\nu_A$ |
| $Re_A$ | 554 | 616 | 353 | 370 | 410 | 426 |
| $T_E$ | 398.5 | 398.8 | 365.5 | 310.7 | 312.6 | 309.0 |
| $\nu_E/\nu_{ref}$ | 1.660 | 1.662 | 1.430 | 1.076 | 1.088 | 1.066 |
| $Re_E$ | 334 | 371 | 247 | 344 | 377 | 400 |

In Section V.A and the Appendix, we have demonstrated that the increased flow temperature within the flame gap could cause reduced vorticity generation in the downstream, although this has limited impact on the upstream-interaction between the two flames. In this part, we have shown that the heat effect could still affect the mode transition indirectly via temperature-dependent fluid properties, such as density and viscosity. Last, our model has certain limitations as it is based on each individual pool flame being flickering and laminar. This means that the pool diameter should vary in a range approximately between $10^{-2}$ and $10^{-1}$ m [15,62], so that the dimension of the flame is large enough to flicker but small enough to not develop turbulence in the near field. The frequency prediction of the dual-flame system is also beyond the scope of this work.

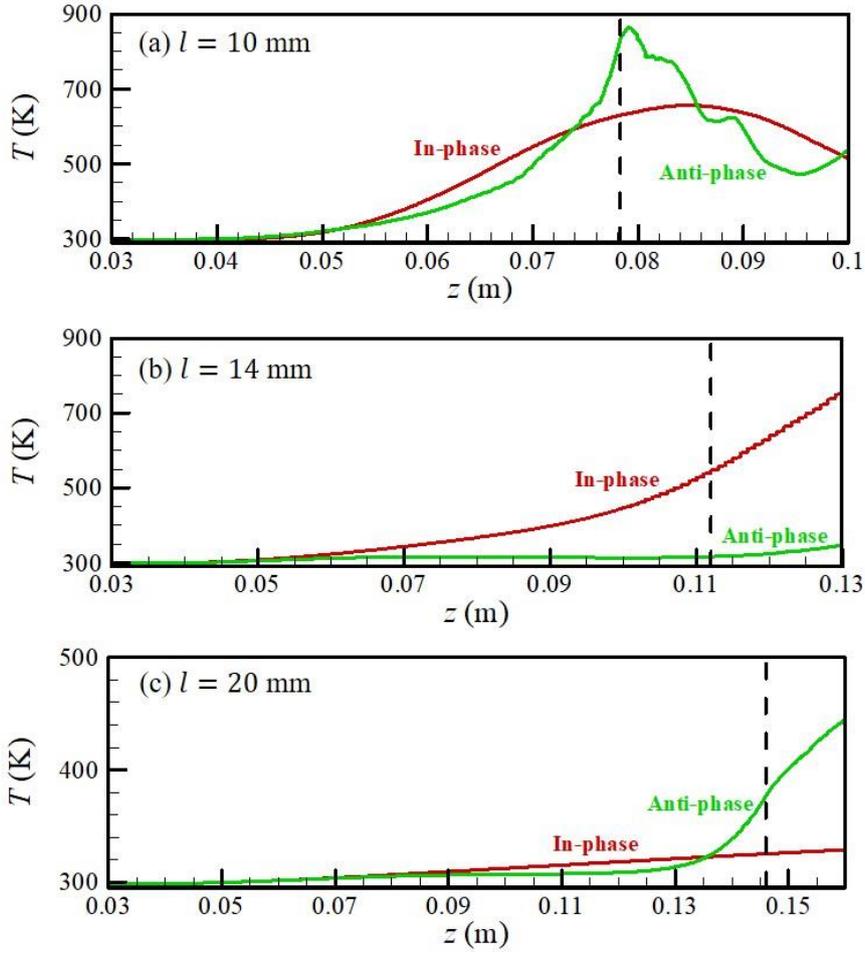

FIG.12. The time-averaged temperature vs. height along the white lines in Fig. 11 for different in-phase and anti-phase flame modes at different gap distances. The dashed lines correspond to the flame pinch-off locations, $z_{po}$.

## VII. CONCLUDING REMARKS

The flickering of dual pool flames was studied numerically and theoretically to understand the dynamic flame behaviors. The anti-phase and in-phase synchronized flickering phenomena visualized by the experiments of Kitahata *et al.* [25] and Forrester [27], respectively, were successfully captured by the present simulation. The heat and flow field information obtained from this simulation provide an insight into the correlation between vortex dynamics and flame flickering. Specifically, the periodic pinch-off of each individual flame in the dual flame system is induced by the formation, evolution, and detachment of the toroidal vortex, which are also known as the mechanism for a single flickering flame. The interaction of the two flames leading to distinct flickering modes was attributed to the coupling effect between the inner-side shear layers of the two flames. Furthermore, the transition between the anti-phase and in-phase flickering modes can be dictated by a characteristic Reynolds number, $Re_A$, which is defined based on the properties of the inner-side shear layers of the two flames. This finding was verified by the collapse of the non-dimensional flickering frequency as a function of $Re_A$, with data from the current simulation and the experiments of Kitahata *et al.* and Nakamura *et al.* [25,26]. Considering the similarities in both the configuration of the vortical structures and the crucial effect of viscous diffusion, the transition mechanism for the different flickering modes is analogous to the transition of the shear layers in a wake flow to form a von Karman vortex street.

The understanding of the current study is not limited to the buoyancy-induced dual-flame system, as the analysis based on the interaction between two interacting shear layers may be extended to the interactions within a multiple-flame system. For example, a planar tri-flame system might be regarded as the interaction of two parallel "von Karman" type of shear layers to yield more flickering modes, in analogy to the double parallel vortex streets in the wake of double bluff bodies [63-65]. In addition, the current work, although focusing on the laminar flames, could shed a light on the study of turbulent flames. Particularly, fires in nature are usually large-scale and turbulent, and the interactions of multiple fires are more destructive and uncontrollable [5-9]. In such turbulent flames, the flame height is determined by a crucial factor, the air entrainment, which is the outcome of the interactions of multiple flames and vortices, and thus is highly unsteady. In this sense, the current work is an elementary one which contributes to understanding the more complex problem of turbulent flames.

## ACKNOWLEDGEMENTS

This work was supported partly by National Natural Science Foundation of China (Grant No. 91641105) and partly by the Hong Kong Polytechnic University (G-UA2M and G-YBGA). The authors are grateful for an additional support by a collaborative open fund from the Key Laboratory of High-temperature Gas Dynamics, Chinese Academy of Sciences.## APPENDIX

This appendix discusses the relative importance between the viscous diffusion effect and the heat diffusion effect in reducing the vorticity of the inner shear layers of two adjacent flames. It has been concluded from the discussion in Section V.A that for a single flame the main contributor to vorticity generation is the buoyancy term (term D in Eq. (6)). The justification for neglecting the viscous diffusion effect (term E) is that the diffusion mechanism only grows the thickness of a standalone vortex sheet while not affecting its total vorticity (circulation). This is however not the same for a vortex sheet which is in close contact with another vortex sheet of the opposite-sign vorticity. In this case, the vorticity diffusion between a pair of counter rotating vortices could be quite strong owing to the large vorticity gradient and would cause a rapid drainage of vorticity from each vortex, known as the vorticity annihilation. Physically, it can be imagined that the annihilation mechanism becomes prominent when the distance between the two vortex sheets are in the same order of or less than the local thickness of the otherwise undisturbed vortex sheet, which is approximately $\sqrt{\nu \Delta t}$ where $\nu$ is kinematic viscosity and $\Delta t$ is the characteristic flickering period. In the following, we further provide a brief demonstration for this argument.

Fig. A1(a) below shows the configuration of a flame-induced vortex sheet. For simplicity, we assume that the vortex sheet is in the vertical direction. According to our previous study [24], the rate of generation of total vorticity (circulation) on a vortex sheet segment of height $\Delta z$ caused by the buoyancy effect can be calculated as

$$\frac{\mathrm{d}\Gamma}{\mathrm{d}t} = \oint \frac{\mathbf{D}\mathbf{u}}{\mathbf{D}t} \cdot \mathrm{d}\mathbf{s} = \oint -\frac{\rho_2 - \rho}{\rho} \mathbf{g} \cdot \hat{\mathbf{s}} \mathrm{d}s = -g\Delta z\left(\frac{\rho_2}{\rho_1} - 1\right). \tag{A1}$$

Since $\Gamma$ also satisfies $\Gamma = \iint \omega \mathrm{d}n \mathrm{d}s$, we can obtain a rough estimation of $\omega$ on the flame side as $\omega_f \approx -(\int 2g(\rho_2/\rho_1 - 1)\,\mathrm{d}t)/d_\mathrm{v}$, where $d_\mathrm{v}$ is the thickness of the vortex sheet, assuming a linear gradient of $\omega$ in the $\hat{\mathbf{n}}$ direction. On the other hand, we can calculate $\mathrm{d}\Gamma/\mathrm{d}t$ for a coupled vortex sheet in Fig. A1(b) as

$$\frac{\mathrm{d}\Gamma}{\mathrm{d}t} = \oint \frac{\mathbf{D}\mathbf{u}}{\mathbf{D}t} \cdot \mathrm{d}\mathbf{s} = -g\Delta z\left(\frac{\rho_2}{\rho_1} - 1\right) + \nu \oint \boldsymbol{\nabla}^2 \mathbf{u} \cdot \hat{\mathbf{s}} \mathrm{d}s = -g\Delta z\left(\frac{\rho_2}{\rho_1} - 1\right) + \nu \frac{\partial \omega}{\partial n}\Delta z, \tag{A2}$$

where we have retained the viscous diffusion term at the vorticity annihilation plane. It is noted that the derivation of Eq. (A2) involves the identity, $\boldsymbol{\nabla}^2 \mathbf{u} = -\boldsymbol{\nabla} \times (\boldsymbol{\nabla} \times \mathbf{u}) + \boldsymbol{\nabla}(\boldsymbol{\nabla} \cdot \mathbf{u})$. Now, the vorticity on the flame side grown after an entire period $\Delta t$ is derived to be $\omega_f \approx -2g(\rho_2/\rho_1 - 1)\Delta t/d_v$, based on which we attain an estimation of the vorticity gradient, $\partial\omega/\partial n \approx (0 - \omega_f)/d_v \approx 2g(\rho_2/\rho_1 - 1)\Delta t/d_v^2$. Plugging $\partial\omega/\partial n$ into Eq. (A2), we have

$$\frac{d\Gamma}{dt} = -g\Delta z\left(\frac{\rho_2}{\rho_1} - 1\right)\left(1 - \frac{2\nu\Delta t}{d_v^2}\right). \tag{A3}$$

We can see that the condition for the viscous term being negligible is $d_v \gg \sqrt{2\nu\Delta t}$; otherwise, the vorticity annihilation would be significant, which is consistent with our physical understanding.

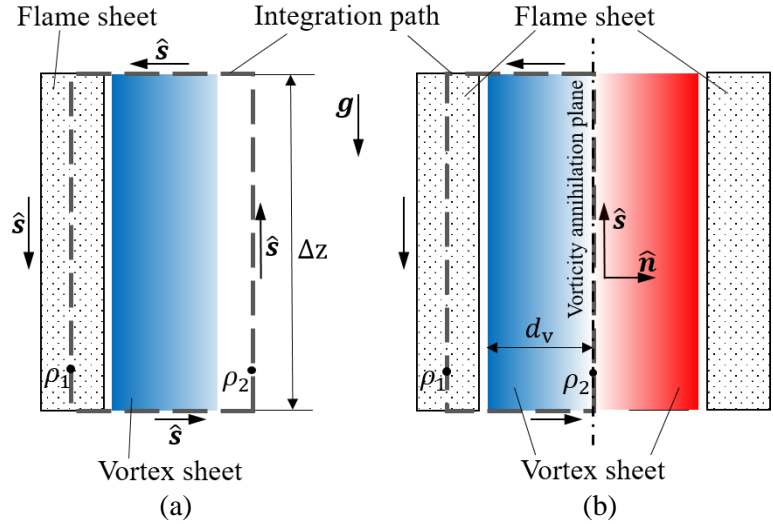

FIG. A1. The vortex sheet configurations for (a) a single flame and (b) two adjacent flames of opposite vorticity.

Furthermore, Eq. (A3) tells that the vorticity generation associated with the temperature difference is proportional to $(T_1/T_2 - 1)$, which is always a positive value since the flame temperature $T_1$ is generally greater than the ambient temperature $T_2$. Thus, the heat diffusion mechanism would at most decrease the generation of vorticity. However, the vorticity annihilation term can be negative and even order-of-magnitude amplified depending on $d_v$. Thus, the viscous diffusion should play a larger role than the heat diffusion effect in reducing vorticity when the two flames are brought together. An example is presented in Fig. A2 below, which shows that the region of significant vorticity diffusion occurs in an upstream location of the rolled-up toroidal vortex, whereas the region of significant heat accumulation is mostly in the downstream of the toroidal vortex. This provides a direct evidence that the vorticity diffusion mechanism happens before the heat effect, and therefore is the primary cause for the reduction of vorticity of the inner shear layers and their subsequent dynamics. Actually, we believe that the extended heat accumulation region observed in Fig. A2(c) is the result of the convective heat transfer induced by the inner shear layers rather than only heat diffusion, as the distance between the inner flames in this region is enlarged by the flame deformation. Nevertheless, the increased temperature of the inner ambient flow could contribute to a slower vorticity generation in the downstream, although it is unlikely a determinant factor for the flickering mode in the upstream.

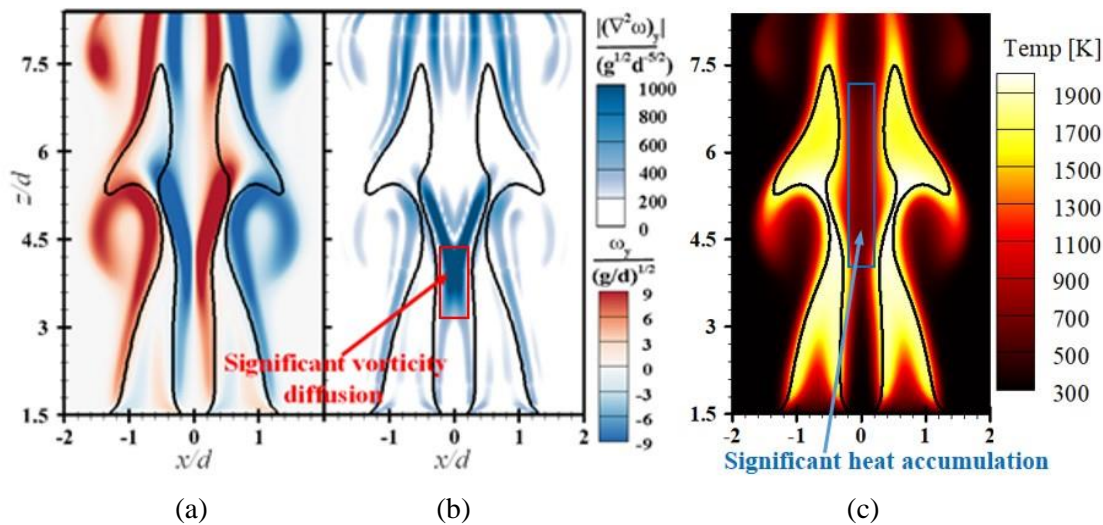

FIG. A2. The contours of (a) vorticity, (b) diffusion of vorticity, and (c) temperature for the in-phase flickering flames of Fig. 6(a).